## This is the preliminary version of the JASIST submission

# Weighted citation: An indicator of an article's prestige

**Erjia Yan[1], Ying Ding**

*School of Library and Information Science, Indiana University, Bloomington, USA*

**Abstract**

We propose using the technique of weighted citation to measure an article's prestige. The technique allocates a different weight to each reference by taking into account the impact of citing journals and citation time intervals. Weighted citation captures prestige, whereas citation counts capture popularity. We compare the value variances for popularity and prestige for articles published in the *Journal of the American Society for Information Science and Technology* from 1998 to 2007, and find that the majority have comparable status.

### 1 Introduction

For several decades, citation counts have played a dominant role in assessing the impact of researchers, journals, institutions, domains, and countries. Along with citation counts, the journal impact factor, proposed by Garfield (1955), is now an established indicator for evaluating the impact of average articles published in a journal. These indicators are easy to calculate and understand. One can consider citation to be a scholarly vote (Davis, 2008). The question is, however, should all votes be counted equally? In citation analysis, citations are unweighted, which means that all citations are given equal weight, regardless of the citing journal and citation time interval.[2] Pinski and Narin pointed out that "it seems more reasonable to give higher weight to a citation from a prestigious journal than to a citation from a peripheral one" (Pinski & Narin 1976, p. 298). Cronin

[1] *Correspondence to*: Erjia Yan, School of Library and Information Science, Indiana University, 1320, E. 10th St., LI011, Bloomington, Indiana, 47405, USA. Email: eyan@indiana.edu

2 Citation time interval is calculated as the date of citation minus the date of publication. For example, two articles published in 2000, one is cited three times in 2001 and the other three times in 2005; the intervals are 1 year and 5 years.

(1984) and Davis (2008) also held that the weight of citations should be differentiated to reflect the prestige of citing journals.

In addition to the prestige of the citing journal, the time of citation also plays an important role in assessing the impact of citations. Different disciplines and journals have different citation half-lives, as reported in the Journal Citation Report (JCR) (Thomson Reuters, 2009). For library and information science, publications at year *n* usually cite publications at year *n*-2 most frequently (see Figure 2), which indicates that a short citation interval is the most common practice, and publications of the previous two years have the most impact on publications at the census year. In their CiteRank model, Walker, Xie, Yan, and Maslov (2007) modified the PageRank algorithm to account for the fact that newer publications have a greater probability of being found through random surfing, by assigning a higher value to these publications. FutureRank (Sayyadi & Getoor, 2009) took a similar approach by adding a personalized vector $R^{time}$ so that more weights were given to recent publications. The AR-index modified the *h*-index by dividing the raw ages of publications (Jin 2007; Jin et al., 2007). The Discounted Cumulated Impact (DCI) Index (Järvelin & Persson, 2008) also takes time into account by using a decay parameter, which devalues old citations.

The status of an actor in a social context is determined by the total number of endorsements received from other actors and the prestige of the endorsing actors (Bollen, Rodriguez, & Van de Sompel, 2006; Ding & Cronin, 2009 submitted). In a similar manner, the status of a journal can be determined by the number of citations received from other journals and the prestige of the citing journals (Franceschet, 2009). Here, we extend this concept to articles, where article status can be defined by three factors: the number of citations the article received, the prestige of the citing journals, and the citation time interval. We define popularity as the number of citations, because all citations are counted equally without consideration of the origin of the citation (e.g., renowned scholars, top journals, etc.), and view prestige as being determined by the weight of citations, which takes into consideration the origin of the citation (i.e., citing journals and citation time intervals).

In this study described herein, we take as our sample articles published in the *Journal of the American Society for Information and Technology* between 1998 and 2007, and weigh each citation on the basis of the prestige of its citing journals and citation time interval. The Article Influence score, similar to the PageRank, is used as the prestige measure for citing journals. An exponential formula is used to evaluate the effect of citation time on articles. The remainder of the paper is organized as follows. In Section 2, we review related work. In Section 3, we present the methodology. In Section 4, we analyze the results and present the findings. Section 5 concludes the study.

## 2 Related work

Currently, weight for citations is calculated at three different levels: author level, journal level, and paper level. At the author level, scholars are interested at assigning different weights to citations based on author self-citation or coauthorship. Contrast to traditional binary counting, Schubert, Glanzel, and Thijs (2005) proposed the fractional self-citation counting. The fractional self-citation counting uses Jaccard Index to determine the overlapping of coauthors between citing articles and citations and gives more weight to the citation if its citing authors are less overlapped with the cited authors. Egghe and Rousseau (1990) discussed how to attribute citations or papers to different contributors. They suggested that the best way is to assign credit proportionally per author for multi-authored papers. Furthermore, they elaborated three methods of distributing weight for authors: straight counting where only the first author's contribution is acknowledged; unit counting where each coauthor's contribution is counted equally; and adjusted counting where each coauthor's contribution is divided based on the number of coauthors.

At the journal level, three approaches are available for the calculation of citation weight: PageRank, SCImago journal rank, and Eigenfactor (Article Influence). Using journal data from the Institute for Scientific Information (ISI), Bollen et al. (2006) used a weighted version of the PageRank algorithm to reflect the prestige of journals. They found that the journal impact factor measures the popularity of journals, whereas PageRank is an indicator of prestige. Leydesdorff (2009) compared PageRank with the *h*-index, impact factor, centrality measures, and SCImago Journal Rank, and found that PageRank is mainly an indicator of size, but has important interactions with centrality measures. SCImago Journal Rank (SJR) indicator, developed by researchers from Spanish universities (SCImago, 2007), applies a PageRank-like indicator to 13,208 journals covered in the Scopus database. SJR is based on the transfer of prestige from one journal to another, and also takes the number of references and number of articles in a journal into consideration (Falagas et al., 2008). Lopez-Illescas et al. (2008) found the correlation between journal impact factor and the SCImago journal rank for journals indexed in 2006 to be very high.

Eigenfactor.org, launched in 2007, is designed to calculate the prestige of scholarly journals (Bergstrom & West, 2008; Bergstrom, West & Wiseman, 2008). The underlying algorithm is based on the idea that a journal is important if it receives many citations from other important journals. The Eigenfactor score is affected by the number of articles in a journal and thus measures the journal's overall importance. As the Article Influence score measures the average influence of papers in a journal, according to Bergstrom and West (2008), it is therefore comparable with the impact factor. Fersht (2009) compared Eigenfactor scores against total number of citations listed in the JCR, and found that there was a strong correlation between them. Davis (2008) found similar results for medical (general and internal) journals.

At the paper level, several studies have applied PageRank algorithms to differentiate the weight of citations. Chen, Xie, Maslov, and Render (2007) used it to assess the relative importance of all publications in the *Physical Review* family of journals from 1893 to 2003. They found that PageRank values and citations for each publication were correlated positively. Ma, Guan, and Zhao (2008) used PageRank to evaluate research impact by country in the field of biochemistry and molecular biology. They also found that citation counts and PageRank were highly correlated. Nevertheless, it should be noted that when the PageRank-like algorithm is applied to paper citation networks, it will amplify the effect of time on such networks (Yan & Ding, 2009 submitted), because the values flow not only to articles directly (via direct citations) but also indirectly (via the citations of citations). As stated by Chen et al., "Long random walks on time-directed networks inevitably drift towards older papers" (2007, p. 14), which suggests that older publications will have an exaggerated value. This differs from journal citation networks, where each node is a journal and each link is the number of citations from a journal to another for the same census year. Citation time, therefore, will not result in amplified PageRank values in journal citation networks. For this reason, we apply the Article Influence score at the journal level and use it to differentiate the prestige of citing journals.

## 3 Methodology

### *3.1 Data in the study*

For this study, we collected all citations to the *Journal of the American Society for Information Science* (*JASIS*) and *Journal of the American Society for Information Science and Technology* (*JASIST*) from 1998 to 2007. We used its official abbreviations: J AM SOC INFORM SCI and J AM SOC INF SCI TEC, and also two major unofficial abbreviations: J AM SOC INFORMATION and J AM SOC INF SCI. Note that we did not exclude author self-citations. Self-citation may affect article rankings (Aksnes, 2003; Glänzel & Thijs, 2004), but we do not investigate it in the present article. Table 1 shows the general statistics of the data under study:

Table 1. Summary statistics of the data*

| | Number |
|---|---|
| Total number of *JASIST* articles (1998-2007) | 1,709 |
| Number of cited *JASIST* articles | 1,476 |
| Ratio of cited articles | 86.37% |
| Total times cited | 8,772 |
| Average number of citations of each *JASIST* article | 5.94 |
| Number of articles that cite *JASIST* | 4,210 |
| Number of cited *JASIST* articles per citing article | 2.08 |
| Number of citing journals that cite *JASIST* | 808 |
| Number of JCR subject areas | 166 |

*cited article: article that is cited by other articles; citing article: article that cites other articles; cited journal: journal that receives citations; citing journal: journal that makes the citation.

*3.2 Methods*

According to the literature, weights for citations can be differentiated in two ways: the prestige of the citing journal and the citation time interval. On this basis, we propose that the prestige of an article is defined by its citing journal and its citation time interval, and that the prestige of a journal is defined by the citing articles and citation time intervals (Figure 1). The assumptions underlying this model are:

- Citations coming from highly-cited journals will have greater prestige than those coming from peripheral journals (Pinski & Narin, 1976; Cronin, 1984; David, 2008); and
- Articles that are cited immediately will have more weight than those being cited at a later date (Zhu, Wang, & Zhu, 2003; Chen et al., 2007; Jin, 2007; Jin et al., 2007; Walker et al., 2007; Järvelin & Persson, 2008; Sayyadi & Getoor, 2009).

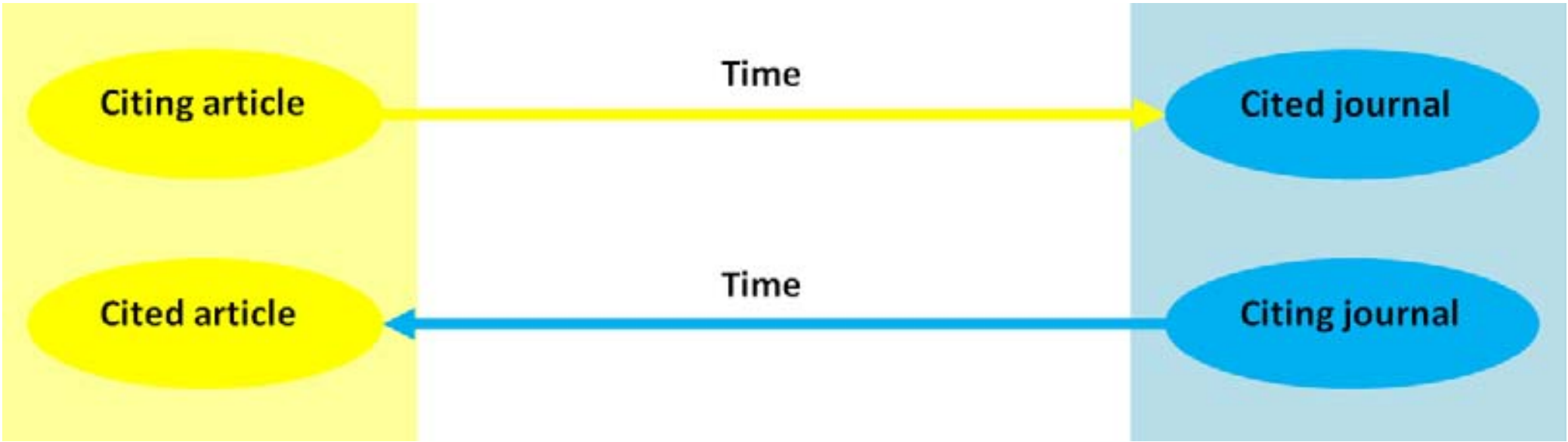

Figure 1. Relationship between article and journal in terms of citation

Similar to Google's PageRank, Eigenfactor (EF) ranks the impacts of academic journals. It uses a five-year target window to calculate citation traffic (Eigenfactor, 2009). Using a window of this length reduces the side-effects of diverse citation patterns caused by different disciplines and permits a fairer evaluation of more theoretical disciplines (Franceschet, 2009). The Article Influence score $AI_{j_i}$ for each journal $j_i$ is a measure of the per-article citation influence of the journal: $AI_{j_i} = 0.01\frac{EF_{j_i}}{\alpha_{j_i}}$, where $EF_{j_i}$ is the Eigenfactor score for journal $j_i$ and $\alpha_{j_i}$ is the number of articles published by $j_i$ over the five-year target window divided by the total number of articles published by all source journals over the same five-year window. Article Influence score is an indicator comparable to the journal impact factor. Eigenfactor scores and Article Influence scores are freely accessible at the Eigenfactor.org, and they have been incorporated into JCR since 2007.

The sum of Eigenfactor scores for all journals in each target period is 1.00. That being so, the Eigenfactor score of journals is comparable within multiple census periods. Meanwhile, $\alpha_{j_i}$ remains stable for different census periods. For the 808 citing journals used in this study, the variance is close to zero ($1.0E-5 \pm 2.0E-6$). Accordingly, the sum of $AI_{j_i}$ may be considered to be the same in each one-year census period. That being so, $AI_{j_i}$ is comparable for different census years.

As noted above, citation time also has an influence on the prestige of articles. At the journal level, citation time is measured by the Immediacy Index, which is a metric for calculating how topical the subject and how quickly a particular journal is picked up and referred to (Elsevier, 2009). Similar to the journal impact factor, the Immediacy Index shows disciplinarity (Yue, Wilson, & Rousseau, 2004). For one field, the Immediacy Index illustrates the prestige of journals (Ray, Berkwits, & Davidoff, 2000). For articles, citation time intervals can also reveal their prestige, where an article that has high prestige will be cited immediately after its publication or even before its formal release (e.g., in preprint format, conference keynote). For example, the two articles described in footnote 2 are both cited three times, which indicates that they have the same popularity. However, the citation time interval for one article is one year but for the other, it is five years. The different intervals suggest that the former article received immediate attention, either because it offered a breakthrough, in which case scholars cite it because it is worthy of trust and endorsement, or because its author or journal has more prestige, which results in scholars in this field paying more attention to the author's works or the journal's publications. Shorter citation time intervals can thus indicate that the cited articles have greater prestige. Figure 2 is based on the number of citations from all JCR journals in 2008 of articles published in *JASIST* in 2008 and backwards (0 in the x-axis indicates 2008, 1 for 2007, etc).

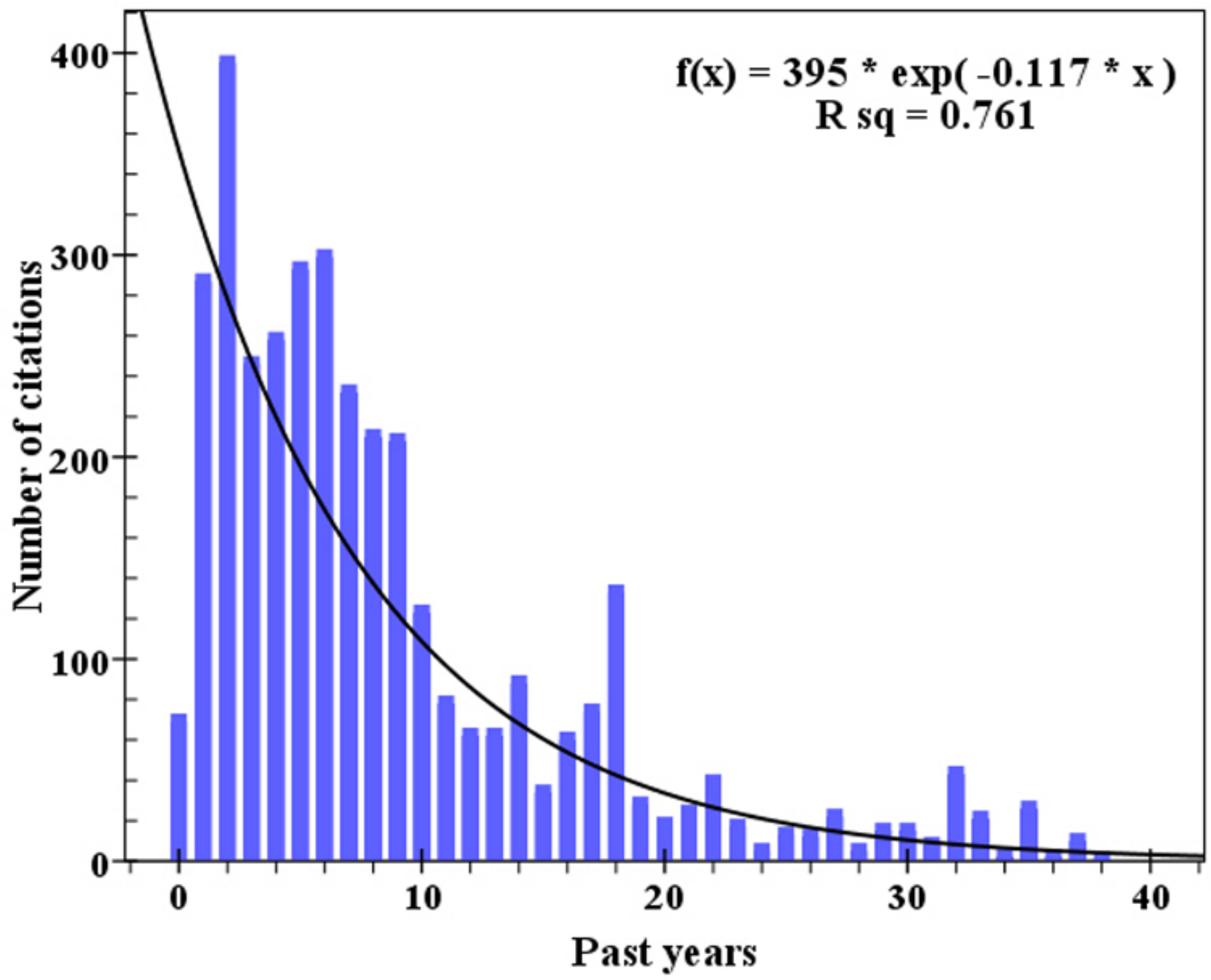

Figure 2. Relationship between number of citations and publication time

Figure 2 shows that the greatest number of citations that an article receives occurs two years after its publication, and that an article's citations in each individual year decrease exponentially (Zhu et al., 2003; Sayyadi & Getoor, 2009). The trend thus fits the curve: $f(x) \sim e^{-0.117x}$. An immediate citation will have greater value, and the cited article in turn is rewarded to a greater degree. Instead of considering each citing article as being of equal weight, we multiply the Article Influence score of its citing journal and the difference between the time of citation and the time of publication $e^{-0.117(t_{citation} - t_{publication})}$, and then sum up the scores for all citing articles, as shown in Figure 3:

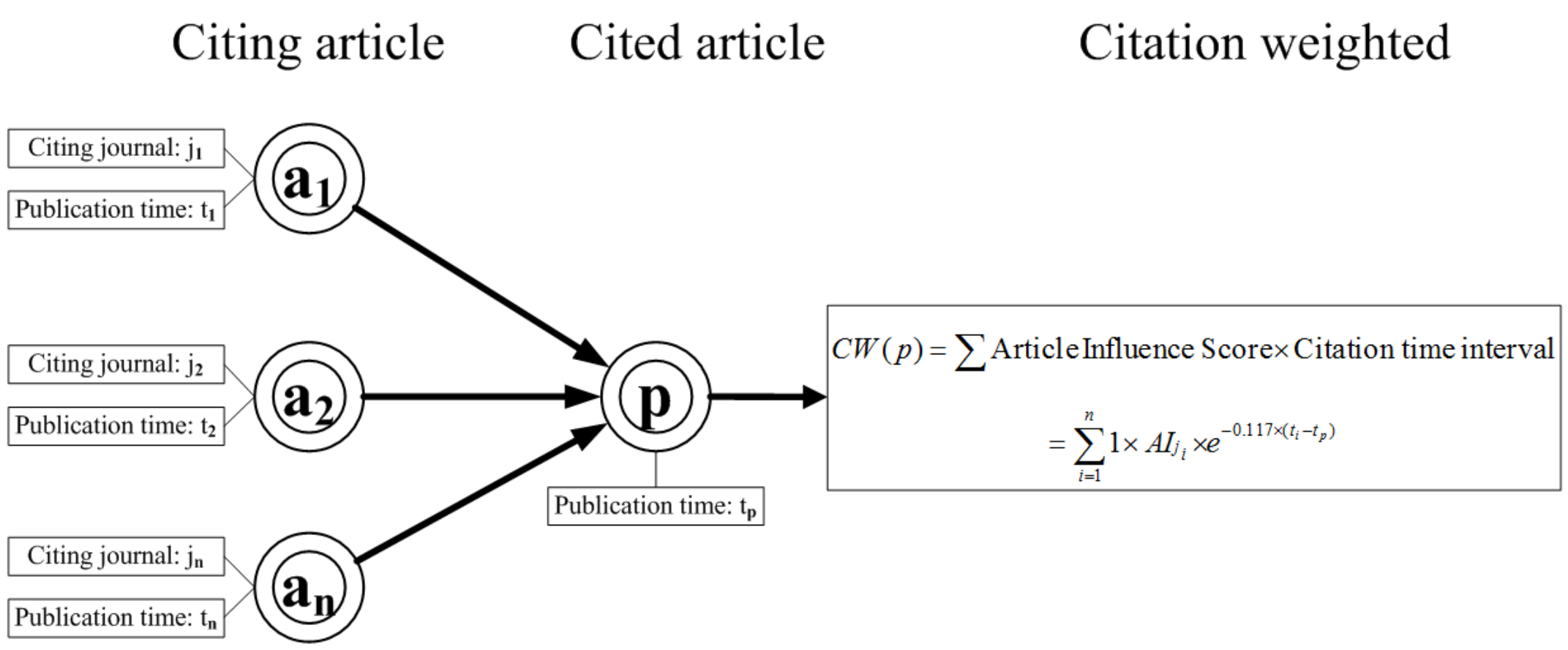

Figure 3. A proposed model for weighted citation

Take a 2005 *JASIST* article as an example: suppose it is cited by article A in 2005, B in 2006, and C in 2007, the following procedure is used:

- First, citing journals were located: articles A, B, and C are published in journal $J_A$, $J_B$, and $J_C$ respectively;
- Second, Article Influence score for each citing journals in each publication year was located: $J_A$ in 2005 has an Article Influence score of $AI_{JA}$, $AI_{JB}$ in 2006 for $J_B$, and $AI_{JC}$ in 2007 for $J_C$, so that the prestige of citing journals are attached to each citation;
- Third, weight for citation time interval was calculated: $e^{-0.117\times(2005-2005)}$ for article A, $e^{-0.117\times(2006-2005)}$ for article B, and $e^{-0.117\times(2007-2005)}$ for article C, so that the citation time is attached to each citation; and
- Finally, multiplication of the Article Influence score by the citation time, and then sum up the product: $e^{-0.117\times(2005-2005)} \times AI_{JA} + e^{-0.117\times(2006-2005)} \times AI_{JB} + e^{-0.117\times(2007-2005)} \times AI_{JC}$ $=1*AI_{JA}+0.89*AI_{JB}+0.79*AI_{JC}$, which is the weighted citation score for this *JASIST* article.

For a simplified application, the third step can be skipped. Therefore, to calculate the weighted citation of a paper, one can 1) locate citations to this paper through Web of Science, Scopus, or Google Scholar; 2) use Eigenfactor.org to obtain Article Influence score for each citing journal (zero if the journal is not covered there); 3) sum citing journals' Article Influence scores, and this will be the weighted citation for this paper. In the similar manner, if we accumulate the weighted citation for an author's publications, the resulting score will be the weighted citation for that author.

## 4 Results

### *4.1 Weighted citation score and citation counts*

Figure 4 shows the scatter plot of weighted citation scores and citation counts. The citation count has a range from 1 to 152, and weighted citation ranges from 0.01 to 76.71. Weighted citation differentiates the prestige of citations: for each citation that comes from a prestigious journal (journal with a high Article Influence score) or one with a short citation time interval, the article cited will have a higher weighted citation score, and vice versa. For each citation count, therefore, the weighted citation has a range of scores, as reflected in Figure 4. The linear regression $R^2$ between weighted citation scores and citation counts is 0.841, which indicates that the two variables have a similar distribution. Several studies confirm this finding, reporting high correlations between citation counts and scores of PageRank-like indicators for journals (Bollen et al., 2006; Davis, 2008; Lopez-Illescas et al., 2008; Fersht, 2009; Leydesdorff; 2009; Bollen et al., 2009; Franceschet; 2009) and also for articles (Chen et al., 2007; Ma et al., 2008; Yan & Ding, 2009 submitted). The high correlation coefficients, however, do not mean that the indicators are interchangeable or that they represent the same information (West, Bergstrom, & Bergstrom, 2009). Although citation counts for a single journal or article may differ extensively from its PageRank values, when considering a large collection of journals or articles, citation counts and PageRank values are always correlated to some extent. In one collection, the majority of journals or articles may have a similar status for citation and PageRank. It is thus not surprising to discover that discrepancies may occur at the local scale that cannot be reflected at the global level. Given this outcome, we compare the value variances for each article in Section 4.3.

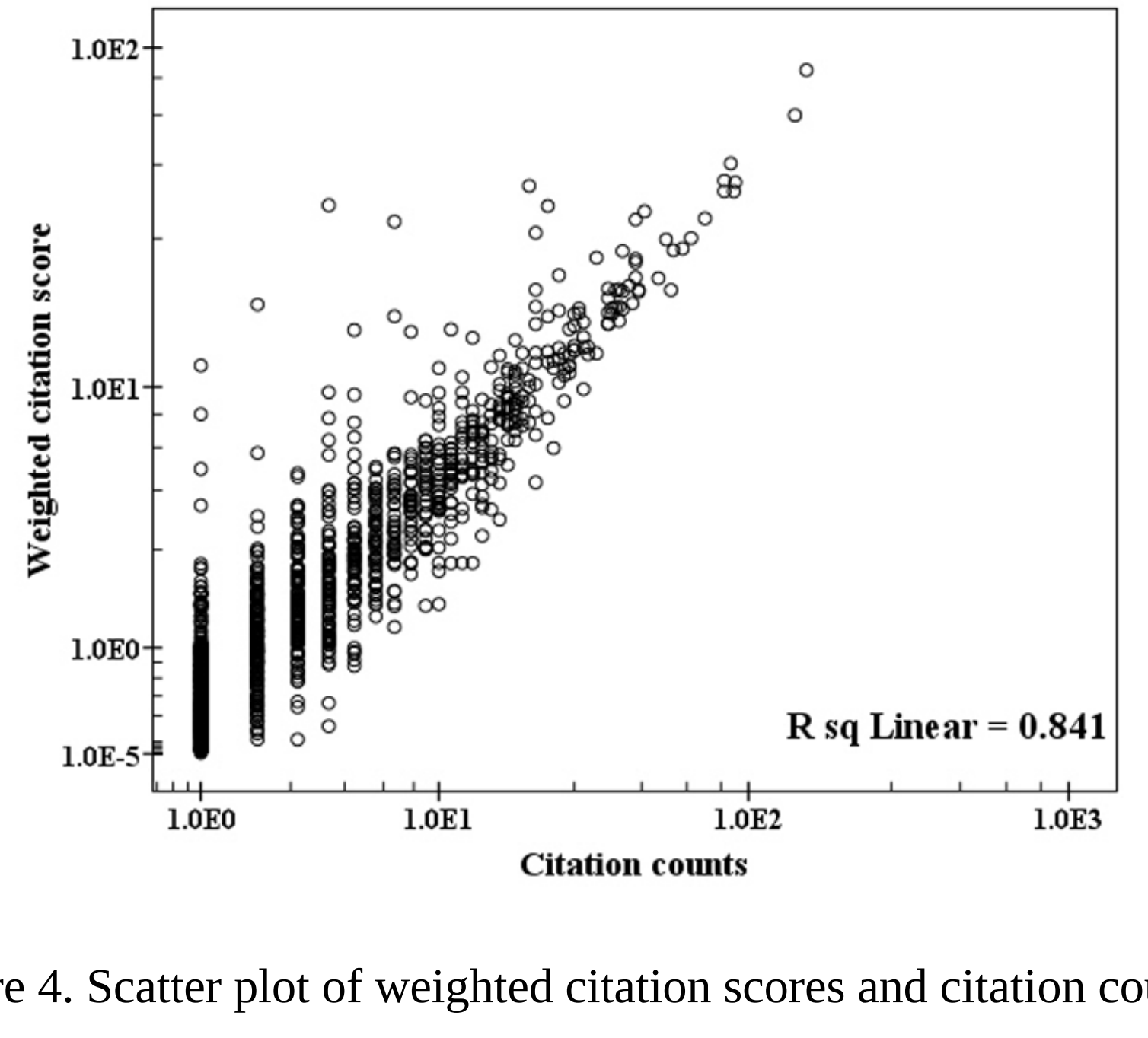

Figure 4. Scatter plot of weighted citation scores and citation counts

Table 2 shows the top 20 articles based on citation counts and weighted citation scores. In this table, 14 articles rank as top 20 for both measures (in bold), and 12 articles appear once in either rank. This result confirms that the majority of articles have equivalent ranking status for popularity and prestige. Six articles' citation ranks outweigh their weighted citation ranks, since some of their citations are coming from journals with lower prestige (i.e. lower Article Influence scores). Meanwhile, six articles' weighted citation ranks outweigh their citation ranks, since some of their citations are coming from journals with higher prestige (i.e. higher Article Influence scores). For example, SRINIVASAN P's 2004 article and van der Eijk CC's 2004 article are cited by several bioinformatics journals which have high prestige; similarly Cronin B's 2001 and 2003 articles and AKSNES DW's 2006 article are cited by *Science* which has quite high prestige.

Table 2. Top 20 articles based on citation counts and weighted citation scores

| Cited References | Citation Counts | CC Rank* | CW Rank* | Cited References | Weighted Citation | CW Rank | CC Rank |
|---|---|---|---|---|---|---|---|
| **SPINK A, 2001, JASIST, V52, P226** | 152 | 1 | 1 | **SPINK A, 2001, JASIST, V52, P226** | 76.71 | 1 | 1 |
| **WHITE HD, 1998, JASIS, V49, P327** | 140 | 2 | 2 | **WHITE HD, 1998, JASIS, V49, P327** | 56.90 | 2 | 2 |
| **SMALL H, 1999, JASIS, V50, P799** | 91 | 3 | 5 | **JANSEN BJ, 2001, JASIST, V52, P235** | 41.27 | 3 | 5 |
| **CHEN HC, 1998, JASIS, V49, P582** | 90 | 4 | 8 | **FIDEL R, 1999, JASIS, V50, P24** | 36.81 | 4 | 6 |
| **JANSEN BJ, 2001, JASIST, V52, P235** | 88 | 5 | 3 | **SMALL H, 1999, JASIS, V50, P799** | 36.40 | 5 | 3 |
| **FIDEL R, 1999, JASIS, V50, P24** | 84 | 6 | 4 | SRINIVASAN P, 2004, JASIST, V55, P396 | 35.55 | 6 | 86 |
| **KLING R, 2000, JASIS, V51, P1306** | 84 | 7 | 7 | **KLING R, 2000, JASIS, V51, P1306** | 34.26 | 7 | 6 |
| **BILAL D, 2000, JASIS, V51, P646** | 73 | 8 | 12 | **CHEN HC, 1998, JASIS, V49, P582** | 34.24 | 8 | 4 |
| **THELWALL M, 2001, JASIST, V52, P1157** | 66 | 9 | 16 | AKSNES DW, 2006, JASIST, V57, P169 | 31.21 | 9 | 479 |
| **SCHACTER J, 1998, JASIS, V49, P840** | 62 | 10 | 18 | CRONIN B, 2001, JASIST, V52, P558 | 31.05 | 10 | 71 |
| **CRONIN B, 1998, JASIS, V49, P1319** | 58 | 11 | 19 | **HAYTHORNTHWAITE C, 1998, JASIS, V49, P1101** | 29.93 | 11 | 15 |
| BATES MJ, 1998, JASIS, V49, P1185 | 57 | 12 | 32 | **BILAL D, 2000, JASIS, V51, P646** | 28.55 | 12 | 8 |
| **LAZONDER AW, 2000, JASIS, V51, P576** | 55 | 13 | 17 | **AHLGREN P, 2003, JASIST, V54, P550** | 28.32 | 13 | 18 |
| HIRSH SG, 1999, JASIS, V50, P1265 | 52 | 14 | 26 | van der Eijk CC, 2004, JASIST, V55, P436 | 27.95 | 14 | 323 |
| **HAYTHORNTHWAITE C, 1998, JASIS, V49, P1101** | 47 | 15 | 11 | CRONIN B, 2003, JASIST, V54, P855 | 25.91 | 15 | 76 |
| BILAL D, 2001, JASIST, V52, P118 | 45 | 16 | 30 | **THELWALL M, 2001, JASIST, V52, P1157** | 25.01 | 16 | 9 |
| WANG PL, 1998, JASIS, V49, P115 | 45 | 17 | 35 | **LAZONDER AW, 2000, JASIS, V51, P576** | 24.77 | 17 | 13 |
| **AHLGREN P, 2003, JASIST, V54, P550** | 44 | 18 | 13 | **SCHACTER J, 1998, JASIS, V49, P840** | 23.28 | 18 | 10 |
| HARTER SP, 1998, JASIS, V49, P507 | 44 | 19 | 22 | **CRONIN B, 1998, JASIS, V49, P1319** | 23.02 | 19 | 11 |
| PALMQUIST RA, 2000, JASIS, V51, P558 | 44 | 20 | 23 | KOEHLER W, 2002, JASIST, V53, P162 | 22.89 | 20 | 24 |

*CW Rank: weighted citation rank; CC Rank: citation count rank

### *4.2 Citation ranking similarity measure*

One way to capture the popularity and prestige of an article is to compare the changes in its rank, an approach implemented by Franceschet (2009). However, this approach is not appropriate for the present study, because citations here follow a discrete distribution: there are only 54 unique ranks and 1,422 duplicate ranks; as a result, many articles share the same rank. By way of contrast, the weighted citation score follows a continuous distribution: there are in total 1,208 unique ranks and only 268 duplicate ranks, so there is a mismatch between the citation rank and the weighted citation rank. As an alternative, we propose our citation ranking similarity measure (CRSM), which compares the discrete value (citation count) and continuous value (weighted citation score). The calculation procedure is as follows:

- Order citation counts and weighted citation scores based on their own ranks;
- For one article, if its citation rank and weighted citation rank are the same, its factor is the ratio between its citation count and weighted citation score; if several articles share the same citation rank, their factor is the ratio of the sum of citation count for this rank to the sum of the weighted citation scores of these articles;
- The intermedium of each article is the product of its weighted citation score and its factor;
- The variance of each article is the subtraction of its citation count from its intermedium. This process is shown in Figure 5 ($CC_n$: citation count for rank n; $CW_n$: weighted citation score for rank n; $F_n$: factor score for rank n; $I_n$: intermedium for rank n; $CR_n$: one *JASIST* cited article; $R_n$: citation rank for this cited article; $R_m$: weighted citation rank for this cited article; and $\Delta$: variance between citation count and intermedium).

| Citation Rank | Citation Count | CW Rank | Weighted Citation | Factor | Intermedium |
|---|---|---|---|---|---|
| n | $CC_n$ | n | $CW_n$ | $F_n=CC_n/CW_n$ | $I_n=CW_n*F_n$ |
| n+1 | $CC_{n+1}$ | n+1 | $CW_{n+1}$ | $F_{n+1}=2*CC_{n+1}/(CW_{n+1}+CW_{n+2})$ | $I_{n+1}=CW_{n+1}*F_{n+1}$ |
| n+1 | $CC_{n+1}$ | n+2 | $CW_{n+2}$ | $F_{n+2}=2*CC_{n+1}/(CW_{n+1}+CW_{n+2})$ | $I_{n+2}=CW_{n+2}*F_{n+2}$ |
| n+3 | $CC_{n+3}$ | n+3 | $CW_{n+3}$ | $F_{n+3}=2*CC_{n+3}/(CW_{n+3}+CW_{n+4})$ | $I_{n+3}=CW_{n+3}*F_{n+3}$ |
| n+3 | $CC_{n+3}$ | n+4 | $CW_{n+4}$ | $F_{n+4}=2*CC_{n+3}/(CW_{n+3}+CW_{n+4})$ | $I_{n+4}=CW_{n+4}*F_{n+4}$ |
| n+5 | $CC_{n+5}$ | n+5 | $CW_{n+5}$ | $F_{n+5}=CC_{n+5}/CW_{n+5}$ | $I_{n+5}=CW_{n+5}*F_{n+5}$ |

⇨

| Cited Ref. | Citation Rank | CW Rank | $\Delta$ |
|---|---|---|---|
| $CR_n$ | $R_n$ | $R_m$ | $CC_{Rn}-I_{Rm}$ |
| $CR_{n+1}$ | $R_{n+1}$ | $R_{m+1}$ | $CC_{Rn+1}-I_{Rm+1}$ |
| $CR_{n+2}$ | $R_{n+2}$ | $R_{m+2}$ | $CC_{Rn+2}-I_{Rm+2}$ |
| $CR_{n+3}$ | $R_{n+3}$ | $R_{m+3}$ | $CC_{Rn+3}-I_{Rm+3}$ |
| $CR_{n+4}$ | $R_{n+4}$ | $R_{m+4}$ | $CC_{Rn+4}-I_{Rm+4}$ |
| $CR_{n+5}$ | $R_{n+5}$ | $R_{m+5}$ | $CC_{Rn+5}-I_{Rm+5}$ |

Figure 5. Calculating the difference between citation count and weighted citation score

Take three cited references in Table 2 as examples: "SPINK A, 2001, JASIST, V52, P226" has $R_n$=1 and $R_m$=1, and its $\Delta=CC_1-I_1=CC_1-CW_1*(CC_1/CW_1)=152-76.71*(152/76.71)=0$; "SMALL H, 1999, JASIS, V50, P799" has $R_n$=3 and $R_m$=5, and its $\Delta=CC_3-I_5=CC_3-CW_5*(CC_5/CW_5)=91-36.40*(88/36.40)=3$; "KLING R, 2000, JASIS, V51, P1306" has $R_n$=6 (two cited references for this citation rank) and $R_m$=7, and its $\Delta=CC_6-I_7=CC_6-CW_7*((CC_{6+} CC_6)/(CW_6+CW_7))=84-34.26*((84+84)/(36.81+34.26))=3.01$. In this way, the difference for articles whose citation ranks and citation weight ranks are the same is zero. For articles sharing the same citation rank, if their weighted citation ranks belong to this citation rank, the difference is close to zero ( $0.00 \pm 0.5050$ ).

*4.3 Popularity and prestige*

Popularity and prestige measure two aspects of an article, and they can be further developed into four cases:

- Low popularity, low prestige: low citation count and low weighted citation score;
- Low popularity, high prestige: low citation count and high weighted citation score;
- High popularity, low prestige: high citation count and low weighted citation score; and
- High popularity, high prestige: high citation count and high weighted citation score.

The majority of articles in a collection have similar popularity and prestige status, being classed as either low popularity with low prestige or high popularity with high prestige. This is illustrated in Figure 6, which shows a strong leptokurtic distribution: the majority of variances are located near zero (standard deviation: 4.2450).

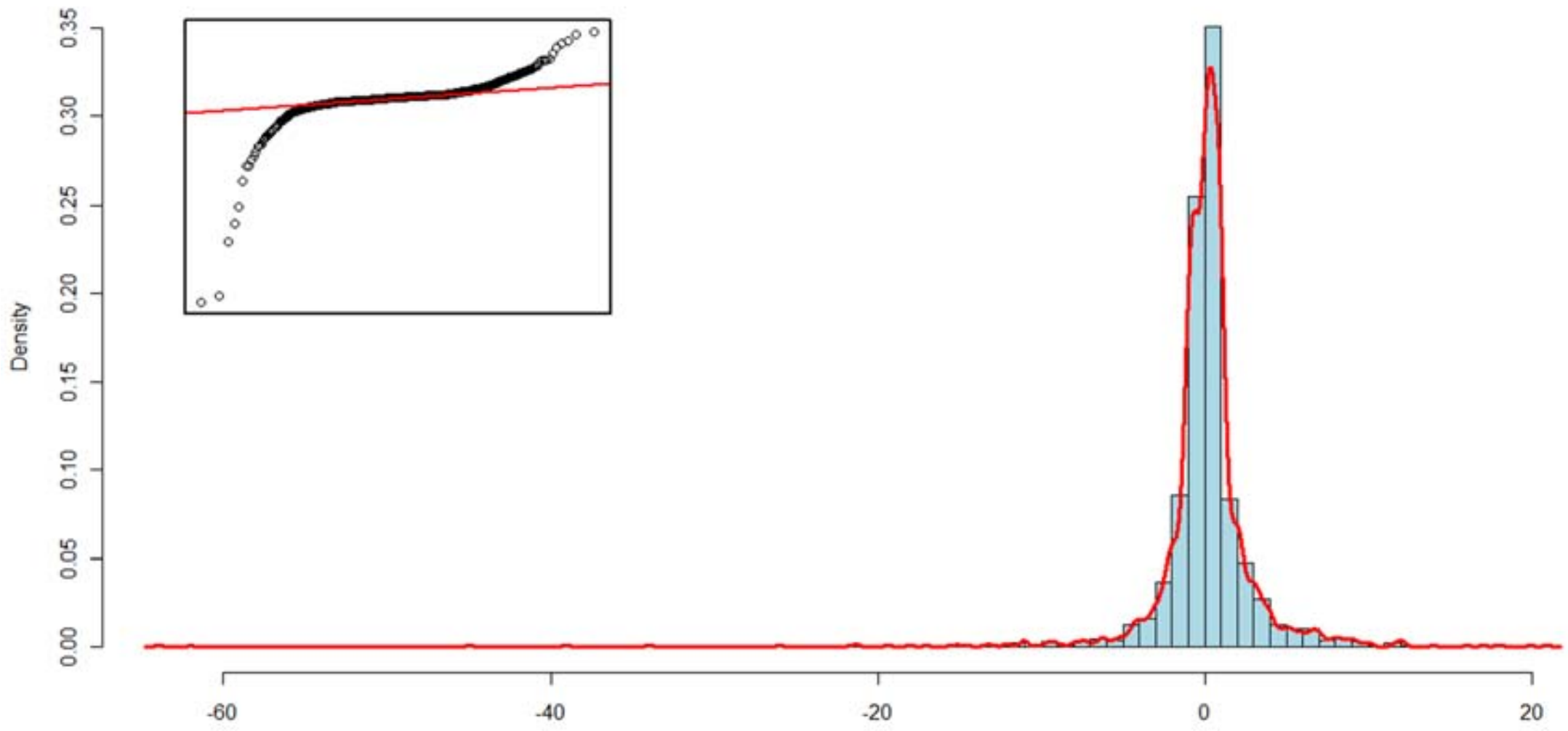

Figure 6. Distribution of variances with quantile-quantile diagram

Table 3 shows the top 20 articles with the greatest (descending) variances between popularity and prestige. They are all oft-cited (minimum citation count: 16), but their citing journal is less prestigious and citation is not immediate. Their popularity thus outweighs their prestige.

Table 3. Top 20 articles with the greatest (descending) variances between popularity and prestige

| Cited References | Citation Counts | Weighted Citation | Intermedium | Δ * |
|---|---|---|---|---|
| THELWALL M, 2001, JASIST, V52, P1157 | 66 | 25.01 | 45.00 | 21.00 |
| BATES MJ, 1998, JASIS, V49, P1185 | 57 | 17.55 | 37.00 | 20.00 |
| SCHACTER J, 1998, JASIS, V49, P840 | 62 | 23.28 | 44.00 | 18.00 |
| CHEN HC, 1998, JASIS, V49, P582 | 90 | 34.24 | 73.00 | 17.00 |
| BILAL D, 2000, JASIS, V51, P646 | 73 | 28.55 | 57.00 | 16.00 |
| CRONIN B, 1998, JASIS, V49, P1319 | 58 | 23.02 | 44.00 | 14.00 |
| SCHWARTZ C, 1998, JASIS, V49, P973 | 30 | 8.75 | 17.75 | 12.25 |
| HIRSH SG, 1999, JASIS, V50, P1265 | 52 | 19.03 | 40.00 | 12.00 |
| DILLON A, 2000, JASIS, V51, P202 | 24 | 5.67 | 12.05 | 11.95 |
| LINGRAS PJ, 1998, JASIS, V49, P415 | 21 | 4.35 | 9.10 | 11.90 |
| WATSON JS, 1998, JASIS, V49, P1024 | 39 | 14.17 | 27.43 | 11.57 |
| ZHANG Y, 2000, JASIS, V51, P57 | 33 | 11.29 | 22.91 | 10.09 |
| LAZONDER AW, 2000, JASIS, V51, P576 | 55 | 24.77 | 45.00 | 10.00 |
| SARACEVIC T, 1999, JASIS, V50, P1051 | 40 | 15.39 | 30.23 | 9.77 |
| BELL DA, 1998, JASIS, V49, P403 | 16 | 3.22 | 6.64 | 9.36 |
| DAVIS PM, 2001, JASIST, V52, P309 | 26 | 8.04 | 16.87 | 9.13 |
| CHEN SY, 2002, JASIST, V53, P3 | 31 | 11.21 | 21.96 | 9.04 |
| BILAL D, 2001, JASIST, V52, P118 | 45 | 17.37 | 36.00 | 9.00 |
| BORLUND P, 2003, JASIST, V54, P913 | 36 | 13.87 | 27.22 | 8.78 |
| LEIGHTON HV, 1999, JASIS, V50, P870 | 36 | 13.94 | 27.37 | 8.63 |

*: Δ =citation count-intermedium

Table 4 shows the top 20 articles with the greatest (ascending) variances between popularity and prestige. Most of them are cited less than 20 times, but their citing journal is prestigious and citation is immediate. Their prestige thus outweighs their popularity. For example, “AKSNES DW, 2006, JASIST, V57, P169” has been cited only four times, but one of its citing journals is *Science*, which has an Article Influence score of 17.353 (as of 2007). As a result, this cited article has a high weighted citation score and displays its prestige in this way. Another example is the two articles written by CRONIN B, where both articles have been cited by a *Science* article, and thus have high prestige.

Table 4. Top 20 articles with the greatest (ascending) variances between popularity and prestige

| CR | Citation Counts | Weighted Citation | Intermedium | Δ * |
|---|---|---|---|---|
| SRINIVASAN P, 2004, JASIST, V55, P396 | 20 | 35.55 | 84.00 | -64.00 |
| AKSNES DW, 2006, JASIST, V57, P169 | 4 | 31.21 | 66.00 | -62.00 |
| van der Eijk CC, 2004, JASIST, V55, P436 | 7 | 27.95 | 52.00 | -45.00 |
| CRONIN B, 2001, JASIST, V52, P558 | 23 | 31.05 | 62.00 | -39.00 |
| CRONIN B, 2003, JASIST, V54, P855 | 21 | 25.91 | 47.00 | -26.00 |
| LEROY G, 2005, JASIST, V56, P457 | 5 | 13.30 | 26.35 | -21.35 |
| BARTLETT JC, 2005, JASIST, V56, P469 | 7 | 14.64 | 28.33 | -21.33 |
| CHEN CM, 2003, JASIST, V54, P435 | 8 | 13.15 | 26.05 | -18.05 |
| WHITE HD, 2003, JASIST, V54, P1250 | 21 | 17.58 | 38.00 | -17.00 |
| MORRIS SA, 2003, JASIST, V54, P413 | 11 | 13.36 | 26.47 | -15.47 |
| LINDSAY RK, 1999, JASIS, V50, P574 | 25 | 19.42 | 40.00 | -15.00 |
| ARTYMIUK PJ, 2005, JASIST, V56, P518 | 4 | 8.57 | 17.37 | -13.37 |
| HUBER JC, 1998, JASIS, V49, P471 | 13 | 12.60 | 26.04 | -13.04 |
| MATIA K, 2005, JASIST, V56, P893 | 5 | 8.41 | 17.06 | -12.06 |
| VAUGHAN L, 2003, JASIST, V54, P1313 | 21 | 15.64 | 33.00 | -12.00 |
| OPPENHEIM C, 2007, JASIST, V58, P297 | 4 | 7.09 | 15.08 | -11.08 |
| AHLGREN P, 2003, JASIST, V54, P550 | 44 | 28.32 | 55.00 | -11.00 |
| HAYTHORNTHWAITE C, 1998, JASIS, V49, P1101 | 47 | 29.93 | 58.00 | -11.00 |
| WEEBER M, 2001, JASIST, V52, P548 | 33 | 21.89 | 44.00 | -11.00 |
| SONG D, 2003, JASIST, V54, P321 | 10 | 10.18 | 19.98 | -9.98 |

*: $\Delta$ =citation count-intermedium

## 5 Conclusion

In the foregoing, we have added a prestige measurement to articles, where the number of citations that an article receives is considered to indicate its popularity and the weighted citation score indicates its prestige. The latter is then defined by two factors, the prestige of citing journals via the Article Influence score and the citation time interval. Comparing the value variances between citation counts and weighted citation score, we find that the majority of articles have similar status for popularity and prestige, which suggests that these articles either have high popularity and high prestige, or low popularity and low prestige. Nevertheless, a portion of the articles have a different status: for low popularity articles that have high prestige, their citing journals are more prestigious and the citation time more immediate, while for high popularity articles that have low prestige, their citing journals are less prestigious and the citation time is less immediate.

The merits of this approach are that the weighted citation score simulates the prestige of an article, it is easy to calculate and understand, and it is practical to implement. In addition, we list the procedures of calculating weighted citation (see 3.2 Methods) and the procedures of comparing weighted citation with citation (see 4.2 Citation ranking similarity measure). Furthermore, the weighted citation can also be extended to author evaluation, including weighted author citation (calculating the accumulative weighted citation for authors) and even weighted author *h*-index (counting weighted citation instead of raw citation).

This study uses the 10-year article data of a single journal and differentiates the prestige of articles on the basis of its citations. In future studies, we will also measure the effect of self-citation by giving less weight to self-citations as studied by Schubert et al. (2005). In addition, we intend to construct an integrated model, wherein the prestige of an article can be defined by three entities: the prestige of its citing authors, the prestige of its citing journal, and the prestige of its citations. Given that the prestige of journals or authors is denoted by the prestige of citations to their publications, values will transfer within articles, journals, and authors via citations until convergence.

**Acknowledgements**

The authors would like to thank Blaise Cronin for his insightful comments on an earlier draft of this paper.